\documentstyle[aps,prl,preprint]{revtex}

\tightenlines
\begin{document}
\draft

\title{Many-body spectrum and particle localization \\
in quantum dots and finite rotating bose condensates}
\author{M. Manninen$^1$, S. Viefers$^2$, M. Koskinen$^1$ 
and S.M. Reimann$^3$}
\address{$^1$Department of Physics, University of Jyv\"askyl\"a,
FIN-40351 Jyv\"askyl\"a, Finland\\
$^2$Department of Applied Physics, Chalmers University of Technology,
S-41296 Gothenburg, Sweden\\
$^3$Mathematical Physics, Lund Institute of Technology,
P.O. Box 118, 22100 Lund, Sweden}
\date{\today}

\maketitle

\begin{abstract}
The yrast spectra (i.e. the lowest states for a given total angular 
momentum) of quantum dots in strong magnetic fields, are studied in
terms of exact numerical diagonalization and analytic trial wave
functions. We argue that certain features (cusps) in the many-body
spectrum can be understood in terms of particle localization due to 
the strong field. A new class of trial wavefunctions supports the
picture of the electrons being localized in Wigner molecule-like 
states consisting of consecutive rings of electrons, 
with low-lying excitations corresponding to rigid rotation
of the outer ring of electrons. The geometry of the Wigner molecule
is independent of interparticle interactions and the statistics
of the particles.
\end{abstract}

\pacs{PACS: 67.40.Db, 73.21.La}

\section{Introduction}
In a sufficiently strong magnetic field, the spin degree of freedom
of electrons in quantum dots is frozen out, and they may be treated
as spinless. Furthermore, if the electron density is low, the strong
field will cause localization of the electrons in a Wigner molecule 
(a finite size realization of a Wigner crystal).
For a circular dot, this localization happens gradually as the magnetic
field increases and is first seen as a separation of a ring of 
electrons\cite{reimann99b,goldmann99}
from the so-called maximum density droplet (MDD)\cite{macdonald93,chamon94}, 
the finite-size analog of the $\nu=1$ quantum Hall \cite{QHbook} 
state with angular 
momentum  $N(N-1)/2$. By increasing the field further,
eventually all the electrons localize\cite{muller96} 
as in a classical one-component
plasma. This localization takes place in the internal coordinate system
of the dot and may be revealed by using a rotating frame or studying
pair correlation functions \cite{maksym96}.

In this paper we study the interplay between particle localization
in a finite 2D electron system in a harmonic confinement 
and the structure of its yrast spectrum,
i.e. its lowest-lying many-body states as function of the total
angular momentum $L$.
Several classes of trial many-body wavefunctions are available
to describe the yrast states of such a system. 
In particular, electron states with angular momenta
equal to odd multiples of $N(N-1)/2$ can be approximately described 
by the Laughlin wavefunction for the 
fractional Hall effect\cite{laughlin83}. Another class of wavefunctions
that has proven successful in 
describing yrast states away from these special 
$L$-values\cite{dev92,jain95}
is due to Jain \cite{jainreview}.

Here, we introduce an additional type of trial wavefunctions which,
although perhaps not as accurate as the Jain construction,
in our opinion give a more intuitive picture of the structure of the
ground state of the dot as one goes from a Laughlin state,
$L=qN(N-1)/2$ to the quasihole- or single vortex state $L=qN(N-1)/2 + N$.
These wavefunctions are similar to those recently introduced by
Bertsch and Papenbrock (BP)\cite{bertsch99} to describe the ground state
of a weakly repulsive, harmonically trapped, rotating 3D Bose condensate 
for $2 \leq L \leq N$.
The close analogy between this system and a fermionic quantum dot
in the quantum Hall regime has been discussed in a number of recent
papers\cite{wilkin98,wilkin00,cooper99,viefers00}; 
the essential observation is that the weak interaction
limit allows for a two-dimensional description of the boson system
in terms of lowest Landau level (LLL) wavefunctions which, just as
for the quantum Hall effect, may contain powers of the Laughlin-Jastrow
factor $\prod_{i<j}^N (z_i-z_j)$, with $z_i$ the complex coordinate
of the $i$th particle. The only difference is the symmetry of the
many-body wavefunction.
This means that it is possible to `borrow' techniques familiar from
quantum Hall physics to describe the rotating Bose condensates and vice
versa.


The outline of the paper is as follows. In section 2 we present
exact numerical calculations for our test case, the six-electron dot.
Trial wavefunctions are introduced and compared to the exact results
in section 3. Finally, we discuss the many-body spectra in terms
of electron localization, using both our numerical and analytical
results.

\section{Exact CI calculations and the yrast spectra}

Independent electrons in a strong magnetic field occupy 
Fock-Darwin states\cite{fock28,darwin30}
with the effective confinement potential
determined by the frequency 
$\omega_{\rm eff}=\sqrt{\omega_0^2+\omega_c^2/4}$,
where $\omega_0$ is the external confinement and $\omega_c$ the cyclotron
frequency $eB/m$. In the absence of interactions, the ground- and
low-lying excited states (for high angular momentum values) are 
entirely in the lowest Landau level
(LLL).

In order to solve the many-body problem in the presence of
Coulomb interactions, we use the configuration interaction 
technique (CI); it has the advantage that
the many-body energy spectrum, not only the ground state, 
can be computed and that
in the case of small particle numbers the solution is 
essentially exact. 
We choose a six electron quantum dot as our test case. 
This is still so small that in the presence of a magnetic field
(the electron system being polarized) 
an essentially exact result can be obtained for several low-energy
many-body states.
The six electron system is the smallest size where the classical 
Wigner molecule has two nearly degenerate isomers
\cite{bolton92}, and as we shall see, the interplay of these
isomers leads to interesting features in the many-body 
spectrum, as first pointed out by Maksym\cite{maksym96}.

In the CI computations we use 
the Darwin-Fock states (with frequency $\omega_{\rm eff}$)
as our single particle basis. For the smallest fields, where the
ground state is the MDD, it was essential to 
include single particle states from several
Landau levels. The Fock states
included were chosen with the help of a cut-off procedure:
The matrix dimension was restricted to about 30000, which
amounts to including about 50 Fock states.
In the high-field case, 
corresponding to filling factor $1/3$, the single 
particle basis was restricted
to the LLL, and a complete CI was used with single particle states
up to $\ell=20$. 
(Note that the $q=3$ Laughlin state, to be discussed below, only 
includes single particle states up to $\ell=15$ for six electrons).

We calculated the many-body spectra for three different values of the 
magnetic field ($b=0.9$, 1.2 and 2.5 in atomic units, corresponding
approximately to filling factors 1, 2/3 and 1/3, respectively). 
The strength of the external confinement was $\omega_0=0.34435$ a.u.
(corresponding to a 2D electron density at the equilibrium density,
 $r_s=1.51$ a.u.).
The many-body spectra are shown in Figs. \ref{fspectra} and \ref{yspectra}. 
For each field strength we show the original spectrum (Fig. \ref{fspectra})
and a spectrum displaying the interaction energy,  
where for each angular momentum value the center of mass
contribution $L\hbar\omega_{\rm eff}$ has been substracted 
(Fig. \ref{yspectra}).

The results show that the overall structure of the spectrum is independent 
of the magnetic field, the role of which is mainly 
to tilt the spectrum so that 
the minimum energy will be at a finite angular momentum\cite{maksym90}.
The spectrum starting from $L_1=15$ is nearly identical for the field
strengths 0.9 and 1.2 and very similar to the spectrum 
corresponding to the strong field $b=2.5$ 
after the angular momentum $L_3=45$. 


\section{Trial wavefunctions}
In order to get some understanding of the microscopic structure
of the yrast states, it is useful to study them in terms of
trial wavefunctions.

At angular momenta $L_q \equiv qN(N-1)/2$ (with $q$ an odd integer), 
corresponding to filling
fraction $\nu=1/q$, an obvious guess for the ground state is the
Laughlin wavefunction\cite{laughlin83}, 
\begin{equation}
\psi_L(q)=\prod_{i<j}^N (z_i-z_j)^q e^{-q\sum_k \vert z_k\vert^2/4},
\label{laughlin}
\end{equation}
where $z_k=\sqrt{m\omega_{\rm eff}/q}(x_k+iy_k)$ are the complex
particle coordinates.

For $q=1$ this wavefunction is a Slater determinant
of LLL single particle states,
$\eta_{\ell} \propto z^{\ell}\exp(-qz_k^2/4)$, 
corresponding to angular momenta from
$\ell=0$ to $\ell=N-1$. 
It has been shown that the wavefunction (\ref{laughlin}) describes
very well the exact many-body state when the angular momentum
corresponds to $q=3$ or larger 
$q$-values, both in a spherical geometry (which mimicks a homogeneous 
state)\cite{haldane85} and for a disk geometry with ``open'' boundary
conditions\cite{laughlin83,trugman85}. 
In the case of $q=1$ the 
Laughlin wavefunction reduces to the (restricted)
Hartree-Fock approximation which, as we shall see, is not a 
quantitatively exact description to the state for $N=6$, 
as the latter contains significant contributions from higher 
Landau levels.

For angular momenta above those described by the Laughlin wavefunction,
$L = L_q + M$, ansatz wavefunctions can be constructed by
multiplying the Laughlin-Jastrow factor by a fully symmetric
polynomial in the $z_k$ of order $M$. In this way,
the good correlations of the Jastrow factor are still built in.
For $M = 1$, the only way of doing this is multiplying
by $\sum z_k$. This is just a center-of-mass excitation of the
Laughlin state. (In a harmonic confinement the center of mass
motion separates from the internal motion.)
For
$2 \leq M \leq N$
we propose
\begin{equation} 
\psi_{LBP}(L_q+M)={\cal S} \prod_{i}^M (z_i-z_0) \, \psi_L(q)
\label{bertsch}
\end{equation}
where the operator $\cal S$ symmetrizes the product, and 
$z_0 \equiv \sum z_k / N$ 
denotes the center of mass coordinate.

The first, symmetric part of this wavefunction was introduced
by Bertsch and Papenbrock\cite{bertsch99} to describe the ground
state of a weakly interacting, rotating Bose condensate for
angular momentum up to the ``unit vortex'' state, i.e. $L \leq N$.
In their case of a delta function interaction, the Bertsch-Papenbrock
wavefunction has been proven to be exact\cite{jackson00,smith00,papen00}.
Here, we combine it with the Laughlin-Jastrow factor to construct
a (fully antisymmetric) wavefunction that interpolates between the
Laughlin state $L=L_q$ and the ``quasihole'' state $L=L_q+N$.
Note that, indeed, for $M=N$ Eq.(\ref{bertsch}) just takes the form
of the wavefunction proposed by Laughlin\cite{QHbook},
for a quasihole centered at $z_0$.

We have tested the applicability of the Laughlin-Bertsch-Papenbrock (LBP) 
wavefunction (\ref{bertsch}) in the present case of six Coulomb-interacting 
electrons in a harmonic potential.

Table 1 gives the 
calculated overlaps of this wavefunction with the exact 
wavefunction starting from $L_1=15$ and from $L_3=45$.
In the first case we determined only the values for $L_1+5$ and
$L_1+6$ which are related to the localization of electrons 
(see next section). The LBP wavefunction seems to be 
a fairly good approximation also for the fermion case,
especially for $q \geq 3$;
the reason that the overlaps for $L=20, 21$ are relatively
poor, is that the Jastrow factor 
itself is not a very good approximation to the exact solution at 
$L=15$. As discussed previously, this is due to contributions 
from higher Landau levels in the exact wavefunction.
If we perform an exact diagonalization for $q=1$ but restricted to
the LLL, these latter overlaps will be much better (1.00, 0.96 and 0.99
for $L=15$, 20 and 21, respectively).


In Table 1 we also give the most important single particle configuration 
of the CI calculation and its amplitude. The MDD is strongly
dominated by only one configuration while the $L_3$ case is much
more dispersed. 
We have argued that the Laughlin-Bertsch-Papenbrock 
ansatz cannot be accurate after the MDD;
nevertheless, the most important configuration is consistent with
the qualitative picture given by the LBP ansatz: For each
additional quantum of angular momentum, one particle is pushed outwards
from the center of the dot\cite{macdonald93}.

There is another class of trial wavefunctions\cite{jainreview},
based on the idea of mapping strongly interacting electrons (or bosons) 
onto weakly interacting composite fermions by attaching an even (odd)
number of flux quanta to each particle. These so-called Jain
states have proven very successful in describing yrast states in
quantum dots\cite{jain95,rejaei93,beenakker93} as well as
weakly interacting rotating Bose condensates\cite{cooper99,viefers00}.
Especially, the method explains the main features of the yrast spectra
already in the noninteracting case, as illustrated in Fig. \ref{yspectra},
and (in the interacting case) gives
excellent overlaps with exact wavefunctions. However,
the Jain construction involves technical steps (formally constructing
states in higher Landau levels and subsequently projecting down to
the LLL in a not uniquely defined way), which generally make it
hard to visualize the qualitative features of the state.
For this reason we choose to study the LBP ansatz which
might not be as accurate as the Jain construction, but
which in a very
intuitive way ``interpolates'' between the Laughlin state $L_q$
and the `quasihole' state $L_q + N$
and in a transparent way links the many-body spectrum to
Wigner crystallization.

\section{Localization of electrons and edge reconstruction}

The original spectra of Fig. \ref{fspectra}
show energy minima (cusps) at angular momenta 15, 20, 25, 30
($b=0.9$ and 1.2) and at 45, 50, 55, 60 ($b=2.5$). This can be understood 
by localization of electrons in a five-fold ring 
(with one atom at the center).
A rigid rotation of a ring of $N$ electrons can only have angular momenta 
$L=nN$ if $N$ is odd and $L=(n+1/2)N$ if $N$ 
is even ($n$ is an integer)\cite{rotbook}.
Other angular momentum values can only be a result of center of mass
excitations or as a combination of the 
rigid rotation with vibrational modes.
The analysis of rotation-vibration spectra 
was found to be successful also in studying localization of non-polarized
electrons in quantum rings\cite{koskinen01}.

The spectra also show markedly low energies when $L$ is 21, 27, 51,
57, corresponding to a six-fold symmetry. Classically the six-electron
system has also a stable geometry of a six-fold ring\cite{bolton92}, 
which is only 
slightly higher in energy than the five-fold ring. In the 
special points $L=15$ and $L=45$ the quantum state
must be a combination of these two structures.  
If the angular momentum is increased by 5, the solution collapses 
into the five-fold symmetry, and
if it is increased by 6 the solution collapses into the six-fold ring.

Cusps of the type seen here, have previously been 
noticed for a few-electron system and 
discussed in terms of particle localization\cite{maksym90}
and symmetry\cite{ruan95,bao97} as well as 
composite fermions\cite{jain95}. Here we will first use the exact
solutions for the six electron dot to test our qualitatively simple
picture of localized electrons.

Figures \ref{dens45} and \ref{dens15} show the radial densities 
of the CI computations for different values of the angular momentum
at the yrast line. Clearly the densities for $L=20$, 25, 45 and 50
are consistent with the five-fold ring, having one particle in the
center, while the results for
$L=21$, 27, 51 have only one ring of electrons.

It is illustrative to study the electron localization also in terms
of analytic trial wavefunctions. The '$N$-particle correlation function'
is simply defined as the square of the wavefunction
\begin{equation}
\rho(z_1,z_2,\cdots,z_N)=\vert \psi(z_1,z_2,\cdots,z_N)\vert^2.
\label{Ncorr}
\end{equation}
Laughlin\cite{laughlin83} showed that in the case of  
Eq. (\ref{laughlin}) finding the maximum of the 
correlation function is equivalent to finding the equilibrium
electron positions in a classical one-dimensional plasma,
where the electrons interact via a repulsive logarithmic
potential and are confined by a harmonic trap.
In accordance with the above discussion, for six particles and
$L=15$ or 45, there are two local minima corresponding to the
five- and six-fold geometries, respectively (the former being
the lowest).

Moreover, Laughlin showed that the optimal geometry is
independent of the exponent $q$ of the wavefunction.
Scaling the complex coordinates with $q$ as we did in
Eq. (\ref{laughlin}), also the distances between
the localized electrons are independent of $q$.
The Laughlin wave function for even $q$-values describes well 
rotating Bose condensates\cite{cooper99,viefers00}.
Consequently, at large angular momenta 
the statistics of the particles 
do not play any role in the internal correlation of the particles.
In fact, this picture is also fairly independent of the actual
interaction potential, since the Laughlin state is an excellent
approximation for long-range Coulomb interactions as well as
zero-range interactions\cite{QHbook}.

The localization becomes stronger when $q$ increases,
making the Laughlin state gradually more similar to a crystal\cite{QHbook}.
This can be measured by the sharpness of the maximum of 
$\rho$. It is easy to show that around a maximum
correlation 
\begin{equation}
\left({{\partial^2}\over{\partial x_1^2}}+
{{\partial^2}\over{\partial y_1^2}}\right)\rho(z_1,z_2^0,\cdots,
z_N^0)_{\vert_{z_1=z_1^0}}=-2q\rho(z_1^0,z_2^0,\cdots,
z_N^0),
\label{Nderiv}
\end{equation}
where $z_i^0$ are the coordinates of the maximum.
This result explains why the density of the MDD is nearly constant,
while the density corresponding to $q=3$ has sharper maxima
consistent with a configuration of electrons localized in a five-fold
geometry. 

Also,
the exact many-body spectra of Fig. \ref{fspectra}
show clearly that the five-fold ring becomes more 
dominating after $L_3$ than after $L_1$, as it drops below
the six-fold one for all $L > L_3$.

Our analytical ansatz of Eq. (\ref{bertsch}) allows for the study of 
localization of electrons at any angular momentum
after any $L_q$. Figure \ref{geom} shows the positions
of electrons at the correlation maxima of Eq. (\ref{bertsch})
for $L=45$,
47, 48, 49, 50, and 51 ($L=46$ is an uninteresting center of mass
excitation). The five-fold ring at 50 and the six-fold ring
at 51 are reproduced as anticipated. In general,
the outer ring consists of as many electrons as is the 
order of the polynomial multiplying the Laughlin wavefunction. 

Note that the geometries are independent of $q$; for example they
would be the same for bosons at $L$ values ranging from 30
to 36. (Or, in fact, for non-integer $q$, which corresponds to
{\it anyons}, i.e. particles obeying fractional statistics, in the 
LLL\cite{hansson95})

For systems containing a large number of electrons, it is not 
possible to perform exact calculations of the yrast
spectra. However, in the region where the particles localize,
simple arguments can be used to predict 
properties of the yrast spectra. The localization 
in the Laughlin state corresponds to the classical plasma
with logarithmic interactions. The ground state 
positions of the electrons are very similar to those
obtained with $1/r$ interactions, which have been 
analyzed by Schweigert and Peeters\cite{schweigert95}. 
In finite systems 
the resulting Wigner molecules are characterized by
rings of electrons (as seen already in the case of six electrons).
Figure \ref{geom18} shows as an example the 18 electron 
geometry. If the Laughlin state is again
multiplied by a polynomial as in Eq. (\ref{bertsch}),
it follows that especially favourable are cases where
the order of the polynomial equals the number of 
electrons in the outermost ring. This case is demonstrated 
in Fig. \ref{geom18} which shows that when the angular momentum
is increased by 11, i.e. the number of electrons in the outer
ring, the only 
effect is the expansion of this outer ring.

The above analysis suggests that starting from the Laughlin state,
the next low-energy state (cusp) is obtained at angular 
momentum $L_q+n$ where $n$ is the number of electrons in
the outermost ring of the corresponding classical electron system.
In the case of the maximum density droplet this 
corresponds to the expansion of the so-called 
Chamon-Wen\cite{chamon94} edge from
the MDD. Indeed the number of electrons in the 
Chamon-Wen edge as calculated by Reimann {\it et al.}\cite{reimann99b}
for $N=20$ is in agreement with the number 
of classical electrons in the outermost ring.
Increasing the angular momentum further
by the number of electrons in the next ring causes also
that ring to expand, and again an energy minimum is 
expected. 

The present scheme thus provides microscopic trial wavefunctions
describing the edge reconstruction of small quantum Hall droplets.
For the special case of the MDD ($q=1$) a similar trial wavefunction
was recently proposed by Goldmann and Renn\cite{goldmann99}.

The expansion of the Chamon-Wen ring also increases the
electron localization. This can be demonstrated 
by studying the decrease of the $N$-particle correlation
when an electron is moved out from the position of the 
maximum correlation (keeping all other electrons in place).
Figure \ref{ncorre} shows the decrease of the correlation 
as a function of the electron displacement (angular average)
for the six electron dot for the MDD ($L=15$), for the
Chamon-Wen state ($L=20$) and for the $q=3$ Laughlin state
($L=45$) calculated using the LBP ansatz
Eq.(\ref{bertsch}). It is clear that in the Chamon-Wen edge
state the electron localization is increased as compared to the 
MDD. The difference between the $L=45$ and $L=15$ states
in Fig. \ref{ncorre} is in agreement with Eq. (\ref{Nderiv}).

Let us finally comment on the case $L=L_q+N$. Our ansatz (\ref{bertsch})
then has the same form as that proposed by Laughlin\cite{laughlin83}
to describe the fractionally charged quasihole 
excitations of large quantum Hall systems. In these systems, the 
quasihole is the lowest-lying bulk excitation.
The finite size of the quantum dot, with no clear distinction between
bulk and edge, allows also for low energy excitations
where one surface layer (of localized electrons) 
at a time is excited. 
As we have argued, the classical interpretation of these excitatitons 
is that the outer ring is rigidly rotating. The lowest energy excitation
then corresponds to angular momentum increase which 
equals the number of electrons in the outermost ring.
This is already seen in the six electron case where
the low energy excitations after the $L_3=45$ state have
$L=50$, 55 etc in accordance with the 5 electrons in the ring
around the center atom.

\section{Conclusions}

We have demonstrated that the localization of electrons 
in a strong
magnetic field can be seen in the structure of 
the many-body spectrum. The many-body yrast states
following each Laughlin state can be approximated
with simple trial wavefunctions similar to those 
originally suggested by Bertsch and Papenbrock\cite{bertsch99}
for rotating Bose condensates. 

In a quantum dot the internal structure of the 
Laughlin state corresponds to localization of
electrons in consecutive rings. The Laughlin-Bertsch-Papenbrock
wavefunction illustrates that the Chamon-Wen edge reconstruction
can be seen as an expansion of the outermost ring.
Since the Laughlin and LBP wavefunctions are good approximations
for fermions as well as bosons and both for long- and short
range interactions, it seems that the internal geometry of the
localized particles in a high rotational state depends {\it only}
on the number of particles in the harmonic confinement.


\begin{figure}
\caption{
Calculated many-body states as a function of
angular momentum for three different values of 
the external magnetic field. The lowest energy states 
corresponding to a five-fold (six-fold) 
symmetry are connected with a dashed (dotted) line. 
}
\label{fspectra} 
\end{figure}

\begin{figure}
\caption{
Many-body states of Fig. 
\ref{fspectra} but with the 
center of mass excitation energy $L\hbar\omega_{\rm eff}$ substracted. 
The structure of the yrast line predicted by the Jain construction
is shown as dashed lines.
}
\label{yspectra} 
\end{figure}

\begin{figure}
\caption{
Radial density distributions of the exact many-body
states in a six electron quantum dot 
at a magnetic field of $b=2.5$ a.u. 
$L$ is the total angular momentum.
}
\label{dens45} 
\end{figure}

\begin{figure}
\caption{
Radial density distributions of the exact many-body
states in a six electron quantum dot 
at a magnetic field of $b=1.2$ a.u. 
$L$ is the total angular momentum.
}
\label{dens15} 
\end{figure}

\begin{figure}
\caption{
Geometry of the `Wigner molecule' of six electrons
for different values of the angular momentum $L$,
calculated from the trial wavefunction Eq.(\ref{bertsch}).
}
\label{geom} 
\end{figure}

\begin{figure}
\caption{
Positions of electrons in an 18-electron dot at the maximum 
of the square of the wavefunction (\ref{bertsch}) for 
$L=qN(N-1)/2=153q$ (filled circles) and $L=153q + 11$, 11 being the
number of electrons in the outermost ring.
}
\label{geom18} 
\end{figure}

\begin{figure}
\caption{
The decrease of $N$-particle correlation
when one electron is moved out
from the maximum position. The distance is given as 
percentage of the nearest neighbour distance.
Solid line $L=15$, dashed line $L=20$, dotted line 
$L=45$. The dot has six electrons.
}
\label{ncorre} 
\end{figure}

\begin{table}
\caption{
Absolute values of the overlaps between the exact 
wavefunction and the approximate wavefunction, 
Eq. (\ref{bertsch}), for different values of the 
angular momentum $L$. MIC is the most important 
single particle configuration as occupation numbers
of the LLL states starting from $\ell=0$. 
The last column gives the amplitude of the MIC.
}
\label{table1}
\vspace{2pt}
\begin{center}
\begin{tabular}{cccc}
\hline
$L$  & $\vert\langle \psi_{BP}\vert\psi\rangle\vert$ & 
MIC & amplitude of MIC\\
\hline
15 & 0.88 & 1111110000000000 & 0.877   \\
20 & 0.87 & 1011111000000000 & 0.743    \\
21 & 0.90 & 0111111000000000 & 0.833    \\
25 &      & 1001111100000000 & 0.634   \\
27 &      & 0011111100000000 & 0.655   \\
\hline
45 & 0.99 & 1000000111110000  & 0.373 \\
47 & 0.94 & 1000000111100100  & 0.180 \\
48 & 0.96 & 1000000110111000  & 0.184 \\
49 & 0.94 & 1000000101111000  & 0.230 \\
50 & 0.93 & 1000000101110100  & 0.290 \\
51 & 0.96 & 0100000011111000  & 0.257 \\
\hline
\end{tabular}
\end{center}
\end{table}

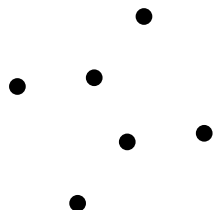
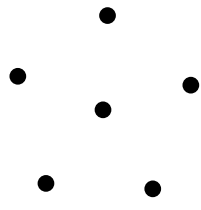
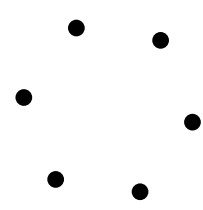
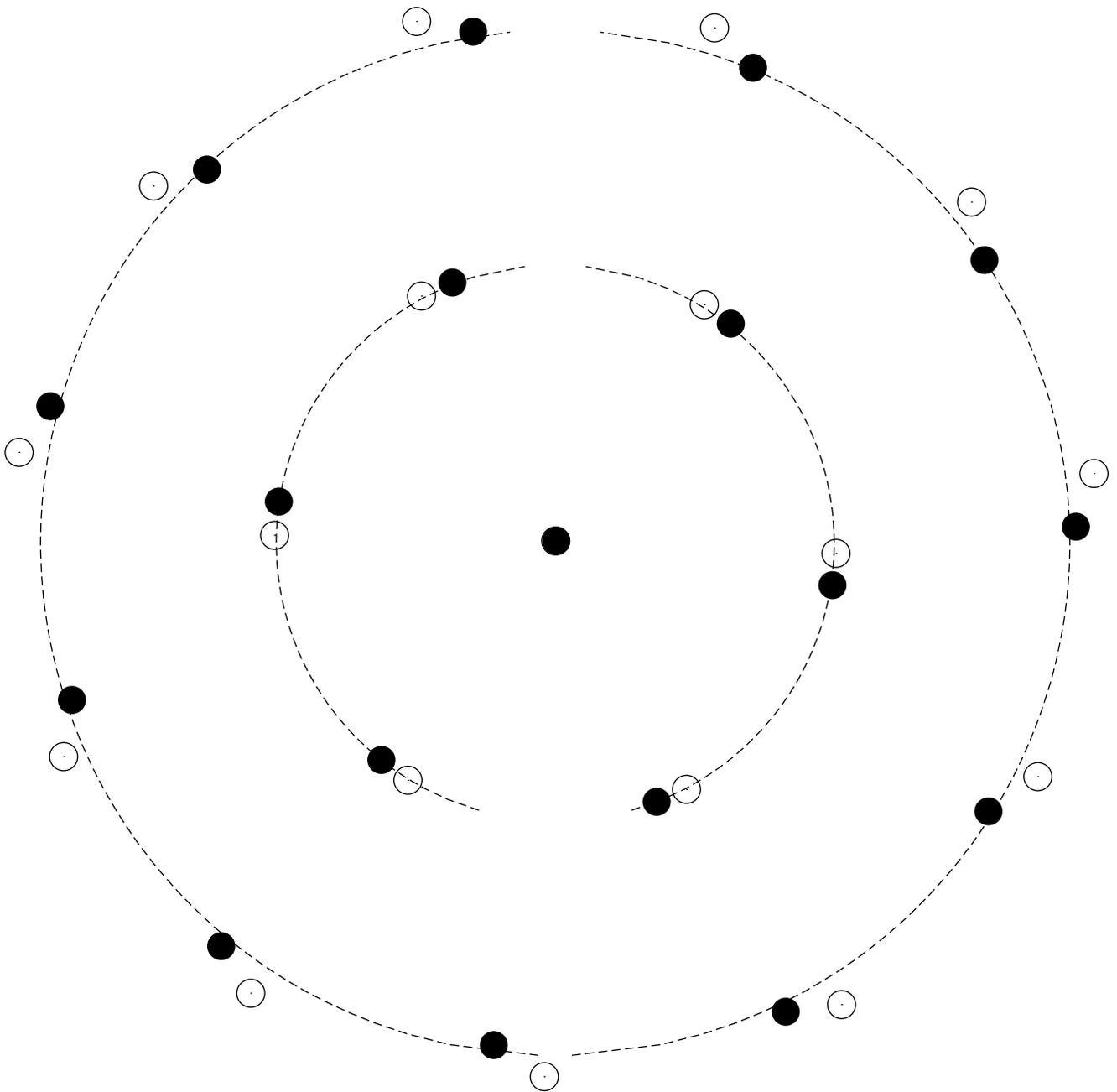
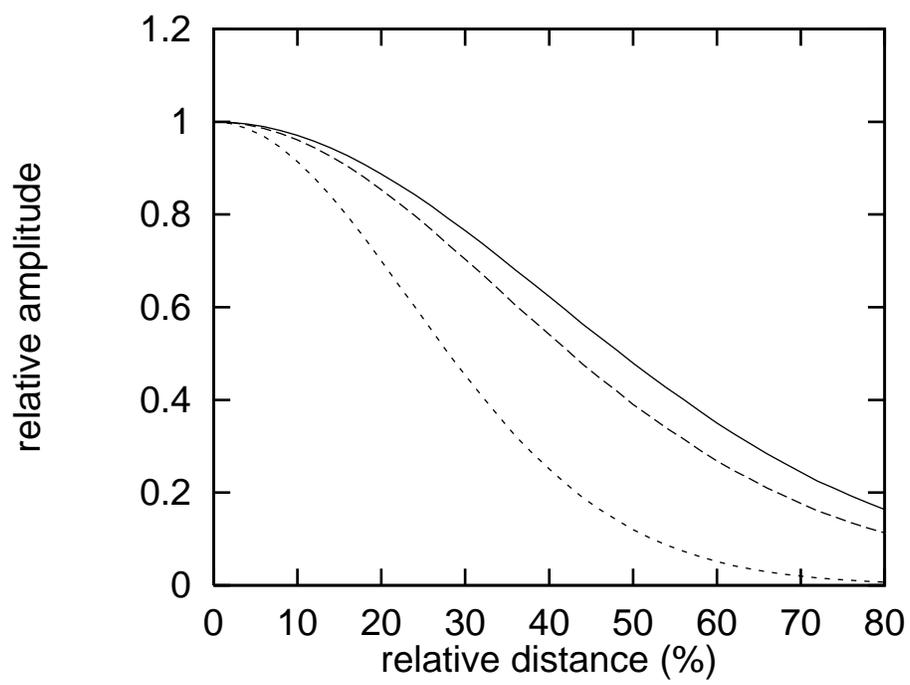
\end{document}